\documentstyle[12pt]{article}
\font\tenbfmi=cmmib10

\def\bdel{\hbox{\tenbfmi\char'016}}

\newcommand{\bc}{\begin{center}}
\newcommand{\ec}{\end{center}}
\newcommand{\bt}{\begin{tabular}}
\newcommand{\et}{\end{tabular}}
\newcommand{\bdes}{\begin{description}}
\newcommand{\edes}{\end{description}}

\newcommand{\be}{\begin{equation}}
\newcommand{\ee}{\end{equation}}
\newcommand{\bea}{\begin{eqnarray}}
\newcommand{\eea}{\end{eqnarray}}
\def\dirac{\delta_D}
\def\half{{{1}\over{2}}}
\def\etal{{\it et al. }}
\title{\bf A Non-Local Formulation of the Peierls Dislocation Model}
\author{Ron Miller and Rob Phillips,\\
Division of Engineering\\Brown University\\Providence, RI 02912, USA\\ \\
Glenn Beltz,\\Department of Mechanical and Environmental Engineering
\\University of California, Santa Barbara\\Santa Barbara, CA 93106-5070, USA\\
 \\and\\Michael Ortiz\\Graduate Aeronautical Labs\\California Institute 
of Technology\\Pasadena, CA 91125, USA}

\begin{document}

\maketitle
\input{psfig}
\section*{Abstract}

Cohesive zone models provide an illuminating and tractable way to
include constitutive nonlinearity into continuum models of defects.
Powerful insights have been gained by studying both dislocations and
cracks using such analyses.  Recent work has shown that as a result of
the locality assumption present in such cohesive zone models,
significant errors can be made in the treatment of defect energies.
This paper aims to construct a non-local version of the
Peierls-Nabarro model in which the atomic level stresses induced at
the slip plane depend in a non-local way on the slip degrees of
freedom.  The non-local interplanar kernel is computed directly from
atomistics and is used to evaluate both the structure and energetics
of planar dislocations.  The non-local formulation does not
significantly change the dislocation core structure from that obtained
with the local model, but the new formulation leads to significant
improvements in the description of dislocation energetics for
dislocations with planar cores.

\section{Introduction}

The appropriate method to use in modeling the behaviour of a material
is often dictated by some intrinsic length scale in the problem.  When
considering atomic scale defects, the nanometer length scale is
dominant, and fully atomistic models are often required.  On the other
hand, macroscopic models are concerned with bulk properties of
specimens with dimensions on the micron scale or larger.  In this
regime, the simplifying assumptions of continuum mechanics are
justified.  Recently, there has been increased interest in modeling
the so-called mesoscale regime, the range of length scales that are
often too large for fully atomistic models but too small for discrete
lattice effects to be ignored.  In this regime, neither atomistic
modeling nor continuum mechanics is satisfactory, and new models which
incorporate features from both approaches seem to be necessary.

One class of model which serves as a bridge between the microscopic
and macroscopic approaches is that of cohesive zone models.  Using a
cohesive zone model allows bulk regions to be handled using
conventional continuum mechanics, while atomistic effects are
incorporated only at certain interfaces where it is deemed that they
are important.  Models of this type were first proposed by Peierls
[\ref{peierls}] to describe dislocations, by Barenblatt [\ref{baren}]
to model fracture processes and by Dugdale [\ref{dugdale}], and Bilby,
Cottrell and Swinden [\ref{bcs}] to estimate plastic zone sizes ahead
of cracks.  A review of a number of examples of this approach may be
found in Miller and Phillips [\ref{miller}], while this paper focuses
on the specific example of the Peierls dislocation.

The Peierls dislocation model has received renewed interest of late,
being used, for example, by Rice [\ref{rice}] in his description of
the brittle versus ductile behaviour of crystals.  The attractiveness
of the Peierls framework is that it offers an analytically tractable
(or at least numerically expedient)
continuum model which incorporates nonlinear features resulting from
the presence of the discrete lattice.  The model assumes that the
atomistic features of a dislocation are confined to a single atomic
plane referred to as the slip plane.  It is only at the slip plane
that discontinuities in the displacement fields are assumed to take
place, and a special constitutive law is used to account for them.
Away from the slip plane, the remainder of the bulk crystal is treated
as a linear elastic medium.  In conventional treatments, the
constitutive law at the slip plane is simplified by assuming that the
energy and stress depend only {\it locally} on the slip distribution,
despite the non-local nature of atomic interactions.  In effect, while
the Peierls model does incorporate a periodic length into the local
relation between stress and slip, it does not define a length scale
below which the non-local effects due to slip gradients become
important.

In earlier work [\ref{miller}] it has been demonstrated that the
failure to incorporate non-local effects is a serious concern, since
full atomistic calculations show that the gradients in the slip
distribution that are present in realistic dislocation cores are often
too large for non-local effects to be ignored.  By postulating a new
form for the energy of a slip distribution, we formulate a model which
includes a non-local term while at the same time reduces to the
original Peierls model in the limit where slip gradients become small.

Many examples exist where a certain model, derived under specific
assumptions, is pushed to the extremes of its range of applicability
and as a result must be corrected through inclusion of higher order
terms.  Anharmonic models of lattice vibrations [\ref{vibes}],
[\ref{vibes2}], recent work on gradient models
[\ref{fleck}],[\ref{aif}], and several papers by Eringen and
co-workers [\ref{eringen1}]-[\ref{eringen3}] on non-local continuum
theories are but a few examples of this strategy.  Similar extensions
are made here in the context of the Peierls framework.  Due to the
inclusion of a new term in the expression for the energy of a slip
distribution, the model is now able to capture non-local effects due
to the presence of gradients in the slip distribution.  This
improvement is demonstrated via comparisons to fully atomistic
calculations of slip plane energies, and is then used to model
dislocation core structures.

In section 2 we begin with a brief description of the original Peierls
framework and its breakdown in the limit of rapidly varying slip
distributions.  In section 3, we present the details of our non-local
formulation and describe how atomistic calculations are used to build
the necessary {\it non-local} constitutive model.  We then use the
non-local Peierls model to compute the energetics of crystalline slip,
and show that the non-local model is in better agreement with purely
atomistic results than was the classical Peierls framework.  Finally,
in section 4, the non-local model is used to obtain the core structure
of a $(100)[011]$ dislocation in fcc Al.

\section{Breakdown of the Traditional Peierls Framework}

In this section, we present a brief explanation of the Peierls model,
but focus mainly on describing the breakdown of the local cohesive
zone assumptions.  For a complete description of the Peierls model,
see Hirth and Lothe [\ref{hirth}].

The Peierls model assumes that a dislocation can be described as two
elastic half-spaces joined at a common plane on which there is a
discontinuous jump in the displacement fields.  We adopt the
convention that the slip plane is the $x-y$ plane, with the dislocation
line along the $y$-axis.  The discontinuity in displacements due to the
presence of the dislocation is referred to as the {\it slip
distribution}, $\bdel(x,y)={\bf u}^+(x,y)-{\bf u}^-(x,y)$, where ${\bf
u}^\pm(x,y)$ are the displacement fields just above and below the slip
plane.  We confine our discussion to the simple case of plane strain
in the $x-z$ plane,
and to the situation where only one component of $\bdel$ is non-zero.
This allows us to write $\bdel$ as a scalar, $\delta(x)$.  Dealing
with the more general case is a straightforward extension of the
results presented below, while the simple case being discussed here is
better suited to demonstrate our arguments.  It is important to bear
in mind that $\delta(x)$ is not constant as in rigid slip, nor is it a
simple step function as in the Volterra model of a dislocation.
Rather, the slip distribution varies from zero at a point on the slip
plane far from the dislocation core to a full Burgers vector once the
core is traversed.  The slip distribution is assumed to lead to atomic
level forces due to the interaction between the slipped surfaces, thus
providing the tractions, $\tau(x)$, on the elastic regions.  These
tractions take a simple form that depends only on the local slip
discontinuity 
\be \tau(x)=\tau[\delta(x)],
\label{tauloc}
\ee 
and can be determined on the basis of atomistic calculations.
Early work within the Peierls framework assumed a simple periodic form
for eqn.~(\ref{tauloc}), with the periodicity tied to the Burgers
vector (see, for example, Foreman, Jaswon and Wood [\ref{foreman}]).
More recently, highly accurate atomistic calculations have allowed for
the direct calculation of $\tau(\delta)$, improving the agreement
between models built upon the Peierls framework and the results of
direct atomistic simulation (see, for example, Rice, Beltz and Sun
[\ref{ricebs}]).

For the purposes of this paper, we find it more convenient to consider
dislocation energetics rather than the resulting forces and tractions.
According to the Peierls model, the energy of a dislocation is made up
of two parts -- the elastic energy contained in bulk regions and the
misfit energy associated with the slip plane.  The first component,
the elastic energy, is fully defined once the elastic constitutive law
for the bulk regions is specified, while the misfit energy is computed
as an integral over the slip plane.  While this is generally a surface
integral, for the simplified geometry of a straight dislocation
oriented along the y-axis, the integral reduces to 
\be
E_L=\int\limits_{-\infty}^{\infty}\Phi[\delta(x)]dx,
\label{misfitloc}
\ee 
where $E_L$ is understood to be the local misfit energy {\it per
unit length} along the dislocation line.  This convention will be
maintained throughout the paper. In eqn.~(\ref{misfitloc}),
$\Phi(\delta)$ is given by 
\be
\Phi(\delta)=\int\limits_{0}^{\delta}\tau(\delta')d\delta'.
\label{phitau}
\ee 
$\Phi(\delta)$ is referred to as the interplanar slip potential,
and can be thought of as the energy cost associated with slipping one
block of atoms over another by an amount $\delta$.  As with the
tractions $\tau(\delta)$, the interplanar slip potential can also be
obtained through simple atomistic calculations (see, for example, Rice
\etal [\ref{ricebs}] or Kaxiras and Duesbery [\ref{kax}]).

The energetic description provided above introduces an important
assumption that has been central to the Peierls framework and results
in a formulation that is strictly {\it local}.  Specifically, this
assumption arises from the fact that the interplanar potential is
computed on the basis of a purely {\it uniform} slip distribution,
even though the actual slip distributions of interest are
non-uniform. The energy of the non-uniform slip distribution is found
using eqn.~(\ref{misfitloc}), which effectively divides the slip
distribution into infinitesimal slip steps, samples the interplanar
potential for each of these steps, and sums the results.  The tacit
assumption of this approach is that despite the {\it non-uniform}
nature of the slip distribution, the local environment at each point
can be considered to be approximately {\it uniform}, allowing for the
local slip energy to be determined from the interplanar slip
potential.  It is expected that as long as the gradients in the slip
distribution are small, this approximation will be valid, but as the
gradients become more severe, the approximation will breakdown.  One
result of interest is to quantify what the maximum acceptable
gradients are, and to determine whether or not the slip gradients
occurring in real dislocation core structures exceed these values.

These questions about the validity of the assumptions in
eqn.~(\ref{misfitloc}) were addressed by Miller and Phillips
[\ref{miller}].  The breakdown of the local assumptions was quantified
by computing the energy of a number of idealized slip distributions in
two ways, first using atomistics (considered ``exact'' in this
context) and then using the approximation embodied in eqn.~
(\ref{misfitloc}).  The parameters in the slip distribution were then
selectively varied, allowing for control of the severity of the slip
gradients and a direct comparison of the two methods of obtaining the
slip plane energy.  These calculations demonstrated a clear breakdown
in the locality assumption.  Further, it was found that the gradients
associated with the slip distributions for simulated core structures
are of the same order of magnitude as those for which
eqn.~(\ref{misfitloc}) failed.

\begin{figure}
\vspace{0.1in}
\centerline{\psfig{figure=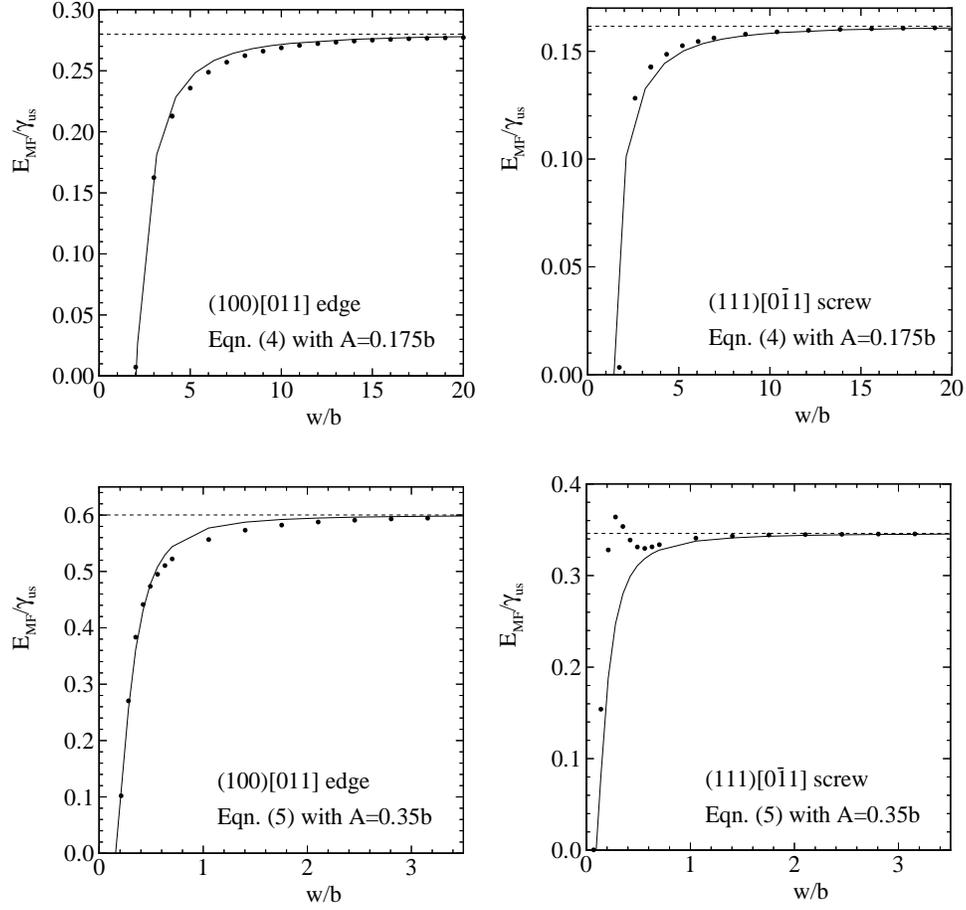,height=5.0truein}}
\caption{Energy of idealized slip distributions as a function of the
parameter $w$, which determines the gradients in the slip distribution.
These results demonstrate the failure of the local slip approximation for
small $w$. Exact atomistic energy (filled circles); 
local model (dashed line); non-local model (solid line).}
\label{singss}
\end{figure}

In fig.~(\ref{singss}), we reproduce the results of Miller and
Phillips for the $\{001\}[110]$ slip system in fcc Al and include new
results for the $\{111\}[110]$ slip system.  The figure shows a plot
of misfit energy per unit area of the slip plane for various slip
distributions.  For periodic distributions, this is computed by
dividing the misfit energy per period by the periodic length.  For
slip distributions which are not periodic, an effective slip area was
used, which is defined as the area over which the slip is greater than
1\% of its maximum value.  The following slip distributions were used,
\be \delta(x)=A\sin {{2 \pi x}\over{w}},\label{sinwav} \ee \be
\delta(x)=A\exp [-\half({{x}\over{w}})^2].\label{gsswav} \ee For these
slip distributions, $w$ can be varied in order to control the slip
gradients.  Small $w$ corresponds to large gradients, and hence the
regime in which we expect the local approximation to fail.  To build
these slip distributions for an atomistic calculation, one divides a
crystal in two, and imposes a different deformation field in the upper
and lower halves.  Judicious choice of these deformations leads to the
appropriate discontinuity at their common surface, and allows for the
computation of the misfit energy due to that discontinuity.  This is
the exact atomistic energy which is compared to the results of
eqn.~(\ref{misfitloc}) in fig.~(\ref{singss}).  Further detail of the
atomistic calculation of misfit energy is given in section 3.

Figure (\ref{singss}) includes two different types of slip
deformation.  In the slip distributions associated with straight
dislocations, two types of gradient effects are possible.  First, both
the direction in which the slip distribution is changing and the
direction of the slip discontinuity itself can be the same.  This is
characteristic of the slip distributions for pure edge dislocations,
and therefore we refer to any slip distribution for which the $\delta$
vector and the $\nabla\delta$ vector are parallel as an ``edge''-type
distribution.  The second type of gradient effect occurs when the
direction of the slip discontinuity is perpendicular to the direction
along which it is changing.  This is characteristic of a pure screw
dislocation, and therefore we refer to such an instance as a
``screw''-type slip distribution.  It is possible to mix these two
effects, but for our present purposes we consider only the pure edge
and pure screw cases.  These two types of slip distributions are
analogous to longitudinal and transverse phonons.

The results in fig.~(\ref{singss}) demonstrate the breakdown of the
Peierls assumption for the slip distributions of eqns.~(\ref{sinwav})
and (\ref{gsswav}) for both the $\{001\}[110]$ edge distribution and
the $\{111\}[110]$ screw distribution.  The plots give the energy of
the slip distributions as a function of the parameter $w$. In all of
the plots, the prediction of the local slip approximation for the
misfit energy is shown as the constant dashed line near the top of
each graph, a striking manifestation of the lack of a
characteristic length scale in the local cohesive zone approach.  
For periodic slip distributions, it is
easy to see why the Peierls model approximation is independent of the
parameter $w$.  The plots of fig.~(\ref{singss}) are of the quantity
\be 
{{E^0_L}\over{w}}={{1}\over{w}}\int\limits_0^w\Phi[A\sin (2\pi
x/w)]dx,
\label{misfitone}
\ee 
where $E^0_L$ is the energy of a single periodic length of the
slip plane.  The fact that this integral correctly represents the
energy of a single periodic length is discussed in section 3.  By
making the change of variables $y=x/w$, one can see that the
expression is independent of $w$.  The data points in
fig.~(\ref{singss}) come from atomistic calculations of the same
misfit energies. Note the divergence from the local estimate at small
values of $w$.

The first mission of the non-local model will be to improve the
agreement between the atomistic and continuum predictions for the slip
energy shown in fig. (\ref{singss}).  The solid line in this figure
shows the improvements made by using the non-local model, and will be
discussed in more detail at the end of section 3.

\section{The Non-Local Formulation}

The notion that any field variable (for example stress, strain or
temperature) is known pointwise, and depends only on other state
variables {\it at that point} is a natural conclusion of observing the
macroscopic behaviour of materials.  The result is the assumption of
locality -- one of the basic tenets of the Peierls model, most other
cohesive zone models, and classical continuum mechanics in general.
However, on the atomic scale the state of each atom is influenced by a
finite cluster of its neighbours, not only by the state at an
individual atomic site.  The non-local formulation of the Peierls
model outlined in this section is proposed as a simple way to include
the non-local nature of atomic interactions into cohesive zone models
such as the Peierls framework.

The modification made to the Peierls model in going from the local to
non-local formulation involves the inclusion of a non-local
contribution to the misfit energy. The addition of this term has
far-reaching implications, though it is the only conceptual change
that is introduced.

Consider again the misfit energy of a dislocation as given in the
local framework of eqn.~(\ref{misfitloc}).  
Non-locality is incorporated by the addition of a term
which should vanish for slowly varying slip distributions while at the
same time capturing gradient effects when they are present.  
For a straight dislocation, we
postulate the non-local misfit energy to be given by 
\be
E_{MF}=\int\limits_{-\infty}^{\infty}\Phi[\delta(x)]dx
+\int\limits_{-\infty}^{\infty}\int\limits_{-\infty}^{\infty}
K(x-x')\delta(x)\delta(x')dx\ dx'.
\label{misfitnloc}
\ee 
The additional term contains the non-local interplanar kernel
$K(x-x')$, which weights the non-local contributions to the total
energy.  On physical grounds, we assume that $K(x-x')=K(x'-x)$, or
that the influence that one point has on another depends only on the
distance between these points.  The new model requires only that we
determine a suitable form for the non-local kernel $K(x-x')$.  We
propose to adopt a similar strategy to that used in obtaining
$\Phi(\delta)$, whereby we extract a numerical reckoning of $K(x-x')$
from atomistic calculations and then fit these results to simple
analytic forms.  In the past, non-local expressions of this type have
been advanced without the benefit of atomistic calculations to
determine the influence function [\ref{eringen1}]-[\ref{eringen3}].
It seems possible that the methods presented here can be adapted to
those cases as well.

Given the non-local kernel, the traction at the slip plane can be
written as the first variation of the misfit energy functional with
respect to the slip distribution, yielding 
\be \tau(x)={{\partial
E_{MF}}\over{\partial \delta}}=
{{\partial\Phi[\delta(x)]}\over{\partial\delta}}
+2\int\limits_{-\infty}^{\infty} K(x-x')\delta(x')\ dx'.
\label{taunloc}
\ee 
Thus, for the non-local model, the traction $\tau$ at any point on
the slip plane depends on the {\it entire} slip distribution, whereas
in the local formulation the stress is determined pointwise.  It is
this additional feature of the new model which allows it to capture
the gradient effects discussed earlier.

To determine the non-local kernel, we solve eqn.~(\ref{misfitnloc})
for $K(x-x')$ by imposing a number of special slip distributions, and
use the non-local term to exactly fit the energies of these slip
distributions to the atomistic result.  This procedure amounts to the
assumption that $K(x-x')$ is independent of the form of the slip
distribution.  While this may not be rigorously true, it is
postulated that the non-local kernel obtained in this way will be
reasonably effective for general slip distributions.  Based on this
assumption, it is convenient to choose slip distributions which allow
for a determination of the Fourier components of $K(x-x')$.  For the
purposes of our calculation we place two demands on the slip
distributions.  First, that it be periodic with period $L$, and
second, that the maximum amplitude of the distribution is small with
respect to the lattice parameter of our material.  The first
requirement is one of computational convenience, allowing for the use
of periodic boundary conditions when determining the exact atomistic
energy of a slip distribution, while the second requirement is made in
order to allow us to rewrite the interplanar potential, $\Phi$, using
its quadratic approximation. This step proves essential in determining
$K(x-x')$ in much the same way that one uses the quadratic
approximation to match the shear modulus in the case of the Frenkel
sinusoid model.  It simplifies the local term in the energy for the
purposes of finding the non-local kernel, but once the kernel is
computed the exact form of the local term will be reinstated.  This
ensures that the non-local model will still be applicable to large
slip deformations.

The energy expression of eqn.~(\ref{misfitnloc}) yields the energy of
the slip distribution over the entire slip plane.  On the other hand,
atomistic models naturally provide us with a way to compute the energy
of only a finite section of the slip plane. If periodic boundary
conditions are used in the atomistic model, as will be used here, then
the energy obtained from atomistics is the energy of a single periodic
length of the slip distribution.  In order to make valid comparisons
between the exact atomistic misfit energy and the energy obtained from
the non-local model, we must find the non-local expression for the
energy of a single periodic length of slip. Equation
(\ref{misfitnloc}) can be re-written in the form 
\be
E_{MF}(\delta)=\sum_{n=-\infty}^{\infty}\Biggl[
\int\limits_{nL+\xi}^{(n+1)L+\xi}c\delta(x)^2dx
+\int\limits_{nL+\xi}^{(n+1)L+\xi}\int\limits_{-\infty}^{\infty}
K(x-x')\delta(x)\delta(x')dx'\ dx \Biggr],
\label{derivea}
\ee 
where the constant $c$ arises from the treatment of the
interplanar potential via its quadratic approximation and $\xi$ is
some origin where $0\le \xi\le L$.  It is easy to show that for
periodic slip distributions, the integrals inside this sum are
independent of $n$ and $\xi$, and hence each periodic length in the
variable $x$ contributes the same amount to the total misfit energy.
We can then write the misfit energy of a single period of the slip
plane, $E_{MF}^0(\delta)$, as 
\be E_{MF}^0(\delta)=
\int\limits_0^Lc\delta(x)^2dx
+\int\limits_0^L\int\limits_{-\infty}^{\infty}
K(x-x')\delta(x)\delta(x')dx'\ dx,
\label{emfo}
\ee 
and note that for a given slip distribution, $E_{MF}^0(\delta)$ is
a quantity which can be computed directly from atomistics.  For this
purpose, it is convenient to take $\delta(x)$ to be \be
\delta(x)=\delta_q(x)=A\sin qx,
\label{derivee}
\ee 
where $q=2\pi/L$ and A is much smaller than the lattice constant
of the crystal under consideration.  We insert this form of
$\delta(x)$ into the expression for $E_{MF}^0$.  Making a change of
variables $z=x-x'$ and recalling that $K(z)=K(-z)$, the energy
expression becomes 
\be E_{MF}^0(\delta_q)=
{{cA^2L}\over{2}}+A^2\int_0^L \sin qx \Biggl[\int_{-\infty}^{\infty}
K(z){{e^{iq(z+x)}-e^{-iq(z+x)}}\over{2i}}dz\Biggr]dx,
\label{deriveb}
\ee 
where we have made use of the exponential form of the sine
function.  We recall the definition of the Fourier transform,
$\hat{f}(k)$, of a function $f(x)$ as 
\be
\hat{f}(k)=\int\limits_{-\infty}^{\infty} f(x)e^{-ikx} dx,
\label{ftrans}
\ee
and also the inverse Fourier transform
\be
f(x)={1\over{2 \pi}}\int\limits_{-\infty}^{\infty}
   \hat{f}(k)e^{ikx} dk,
\label{funtrans}
\ee
and use eqn.~(\ref{ftrans}) to write the non-local energy expression
in terms of the Fourier transform of the non-local kernel,
$\hat{K}(q)$.  Noting again that $K(z)=K(-z)$, we see that
$\hat{K}(q)=\hat{K}(-q)$, and the expression for the total energy becomes
\be
E_{MF}^0(\delta_q)={{cA^2L}\over{2}}+A^2\hat{K}(q)\int_0^L\sin^2 qx dx.
\label{derivec}
\ee
Evaluating the integral leads to an explicit expression for the non-local
interplanar kernel in Fourier space;
\be
\hat{K}(q)={{2E_{MF}^0(\delta_q)}\over{A^2L}} - c.
\label{khat}
\ee
At this point, the strategy is to obtain the energy dependence on the
Fourier variable $q$ numerically through an atomistic model.

\subsection{ Atomistic Determination of the Non-Local Interplanar Kernel}

The procedure for determining the energy of a given slip
distribution was outlined in section 2 and is described in detail by
Miller and Phillips [\ref{miller}].  Here, we use the
same computational approach which is described briefly below.

We desire the energy of the slip distribution given by eqn.~(\ref{derivee}) as 
a function of the parameter $q$.  To make the computation, the crystal
is divided into an upper and a lower half.  By imposing
the appropriate displacement fields on the two half spaces, we can
create the desired slip distribution at their common interface.  For example,
consider the slip distribution of eqn.~(\ref{derivee}).	For the case of
edge-type slip, the slip distribution can be obtained by imposing the 
following displacement fields on the crystal:
\be
{\bf u^+}=[(A/2)\sin qx,0,0],
\label{utop}
\ee
\be
{\bf u^-}=[(-A/2)\sin qx,0,0],
\label{ubot}
\ee
where ${\bf u^+}$ represents the displacement field in the upper half 
of the crystal and ${}\bf u^-$ is that in the lower half.  This combination 
of displacement fields leads to a slip distribution of the form
\be
\delta(x)=u_x^+-u_x^-=A\sin qx
\label{upm}
\ee
at the plane where the upper and lower half spaces meet.

Using any convenient atomistic model (we have used the embedded atom method
(EAM), see [\ref{eam}] for example.)
it is then possible to compute the energy of this crystal.  The
result of such an atomistic calculation is an energy consisting of
two parts, that due to the interface and an additional elastic strain
energy due to the deformation in the bulk regions of the crystal.  The
elastic energy can be found directly by computing the energy of
each half space separately, using a periodic computational cell containing no
slip discontinuities.  This energy can then be subtracted from
the total energy of the configuration which includes the slip
jump, leaving the misfit energy, $E_{MF}^0$.

\begin{figure}
\vspace{0.1in}
\centerline{\psfig{figure=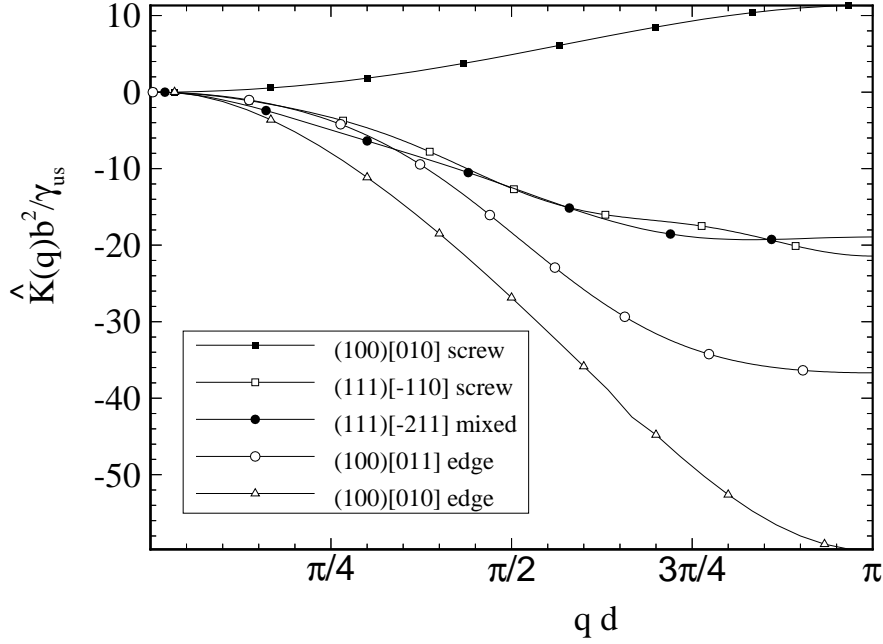,height=3.7truein}}
\caption{Examples of the Fourier space representation of the 
non-local interplanar kernel, $\hat{K}$,
for various slip systems in Al, as
obtained using the EAM potentials of Ercolessi and Adams [19].
See table (1) for normalization constants.}
\label{klibrary}
\end{figure}

By repeating this procedure for a sequence of values of $q$, we obtain
a discrete representation of the non-local kernel $\hat{K}(q)$ in
Fourier space. Results of such calculations using the EAM potentials
for Al of Ercolessi and Adams [\ref{ercadams}] are presented in
fig.~(\ref{klibrary}). The various slip systems are either edge, screw
or ``mixed'', where edge and screw are as defined in section 2.  The
``mixed'' system is in this case the direction associated with the 30
degree Shockley partial.  All of the curves are given in
non-dimensional form using the constants given in table
(\ref{dattab}), where $\gamma_{us}$ is the unstable stacking fault
energy for a given slip system.  The values in the table are obtained
by rigidly sliding two blocks of atoms with respect to one another and
allowing relaxations in only the out-of-plane direction.  The
remaining constants in the table are $b$, the Burgers vector, and $d$,
which is defined below.  The representative examples in
fig.~(\ref{klibrary}) demonstrate a number of the characteristics of
the non-local interplanar kernel when constructed in Fourier space.
All such functions are even and periodic, and therefore only the first
half period of each is shown.  The periodic length of $\hat{K}$
depends on the slip system being considered, and is given by $2\pi/d$
where $d$ is the distance between planes with normal parallel to the
direction of the slip gradient vector.  Note that these planes are a
set of y-z planes, which are perpendicular to the x-y slip plane.
Another characteristic feature of the $\hat{K}$ function is that it is
exactly zero at $q=0$, which implies that in the limit of very slowly
varying slip distributions, the non-local correction vanishes and we
recover the classical Peierls framework.  It is also worth noting that
while most of the $\hat{K}$ functions are negative, we see that for
the case of screw-type slip distributions on the $\{001\}[010]$ slip
system $\hat{K}$ is entirely positive.  This suggests that there is no
set rule about whether the non-local contribution to the misfit energy
is positive or negative -- it will depend on both the slip system
being considered and on the Fourier components of the slip
distribution itself.  Comparing the various curves in this figure can
help us to understand the relative importance of the non-local effect
for various slip systems.  Note that the strongest effect is
associated with the $(100)[010]$ edge system.  It is interesting that
this same slip system, but in the screw orientation, shows the weakest
non-local correction, and that the sign of this correction for the
various Fourier components of the slip distribution is reversed. 
This is because the screw-type sinusoidal slip distribution represents
a more severe misfit configuration than the edge-type for this
slip system.  Meanwhile, the local model predicts the same energy for
either edge or screw slip due to the symmetries of the $\{100\}$
planes.  For this slip system, the local
prediction for misfit energy somewhat overestimates the edge energy,
while somewhat underestimating its screw counterpart.  The result is
this seemingly anomalous behaviour of the screw-type slip. 

\begin{table}
\bc
\bt{|c|c|c|c|}
\hline
Slip System & $b$ (\AA) & $d/b$ & $\gamma_{us}$ ($eV/\AA^2$) \\
\hline
$(100)[010]$ screw & 4.032 & 1/2 &  0.04065\\
$(111)[\bar{1}10]$ screw & 2.851 & $\sqrt{3}/6$ &  0.02705 \\
$(111)[\bar{2}11]$ mixed & 1.646 & $\sqrt{3}/2$ & 0.008032 \\
$(100)[011]$ edge & 2.851 & 1/2 &  0.02521 \\
$(100)[010]$ edge & 4.032 & 1/2 &  0.04065\\
\hline
\et
\ec
\caption{Burgers vector $b$, interplanar spacing $d$ and unstable stacking 
fault energy $\gamma_{us}$ for the slip systems considered in fig. (2) 
as obtained from the EAM Al potentials of Ercolessi and 
Adams [19].}
\label{dattab}
\end{table}

\subsection{ Approximate Analytic Form in Real Space}

\begin{figure}
\vspace{0.1in}
\centerline{\psfig{figure=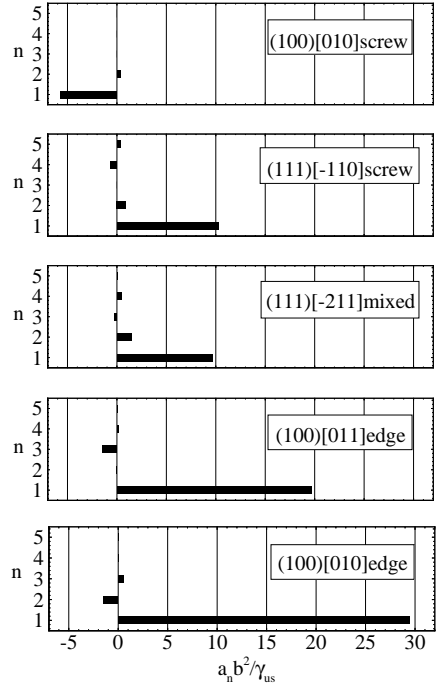,height=3.7truein}}
\caption{Non-local influence coefficients, $a_n$, for the slip systems
considered in fig. (2).}
\label{abars}
\end{figure}

Numerical inversion of the Fourier representation of $\hat{K}$ deduced
above can
be problematic due to the fact that $\hat{K}$ is known only for a discrete
set of points in Fourier space.  Therefore, we propose to fit $\hat{K}$ in
Fourier space with a cosine series, which can be easily transformed 
into real space in the form of a sum of Dirac delta functions.  Because
$K(x-x')$ exists as part of an integrand over the entire slip plane, these
Dirac deltas will have the effect of reducing the dimension of the integral
in eqn.~(\ref{misfitnloc}).  We have found that excellent fits of the numerical
data can be obtained using only the first few terms of such a cosine 
series, and one can therefore write 
\be
\hat{K}(q)\cong {{a_0}\over2} + \sum_{n=1}^N a_n\cos ndq,
\label{khatseries}
\ee
where $a_n$ are the fitting parameters and $d$ is the distance between 
planes perpendicular to the slip plane, as described previously.  The
number of fitting parameters $N$ will depend on the range of the atomistic
potentials used in obtaining $\hat{K}$, but we have found that for 
the potentials and slip systems considered here, the 
value of $N$ typically does not need to exceed five.  This form for the
non-local kernel allows for easy Fourier inversion, yielding
\be
K(x-x')=
{{a_0}\over2}\dirac(x-x')+\sum_{n=1}^N{{a_n}\over2}
\Bigl( \dirac(x-x'+nd) + \dirac(x-x'-nd) \Bigr),
\label{kreal}
\ee
where $\dirac$ is the Dirac delta function.  This
expression for the non-local kernel can be used in
the original definition for the misfit energy, as well as in the expression
for the tractions on the slip plane.  The results are
\be
E_{MF}=\int\limits_{-\infty}^{\infty}\Phi[\delta(x)]dx
  -\sum_{n=1}^N\int\limits_{-\infty}^{\infty}
   a_n\bigl( \delta(x)^2-
   {{\delta(x)\delta(x-nd)+\delta(x)\delta(x+nd)}\over2} \bigr)dx,
\label{misfitappr}
\ee
\be
\tau(x)=
{{\partial\Phi[\delta(x)]}\over{\partial\delta}}
-\sum_{n=1}^Na_n\bigl(2\delta(x)-\delta(x+nd)-\delta(x-nd)\bigr),
\label{tauappr}
\ee
where we have eliminated $a_0$ from the expressions by using the fact that
$\hat{K}(0)=0$.  The $\delta$ appearing in this expression is the slip 
distribution, and not the Dirac delta function $\dirac$.  
The computed values of $a_n$ are plotted in fig. (\ref{abars}) 
for the five slip systems considered here.  In this figure, 
increasing $n$ corresponds to sampling the slip distribution farther and
farther from the point at which the energy and stress is being computed.
It is clear that the non-local effects decays quite
rapidly, and in all cases $a_1$ (corresponding to the near neighbour 
non-local influence) is by far the most important contribution.  Non-local
influence is felt as far as five neighbours away in some systems, but 
beyond $a_5$, the coefficients are negligible.

Note that the expressions for the energy and stress are
consistent with our intuition as to the nature of the
non-local terms.  First, in the limit when
$\delta(x)$ is slowly varying, $\delta(x\pm nd)\rightarrow\delta(x)$ and
the above expressions reduce to the conventional local formulation.
Second, the
non-local effects for a given point arise as a result of sampling a discrete
set of points along the rest of the slip distribution.	These points are
spaced by a distance $d$, which coincides with the spacing of atomic
planes in the direction of the slip gradient vector.  The discreteness
of this sampling, together with the physical significance of the spacing
of the sampling points results in a sensible atomistic extension of the 
original constitutive assumptions of the local model.

Finally, it is interesting to note that the terms of the sum in eqn.
(\ref{tauappr}) can be viewed as a linear combination
of approximations to the derivatives of the slip distribution.  The
symmetry of our non-local kernel in Fourier space means that only 
even order derivatives contribute to this correction, and therefore 
it is possible to re-write eqn. (\ref{tauappr}) in the form
\be
\tau(x)=
{{\partial\Phi[\delta(x)]}\over{\partial\delta}}
-\sum_{n=1}^Nc_{2n} \delta^{(2n)}(x),
\label{tauappr2}
\ee
where the coefficients $c_n$ are the weights associated with the 
$n$th derivative, and $\delta^{(n)}$ is the central difference approximation
to the $n$th derivative of the slip distribution.  The $c_n$ are simply linear
combinations of the $a_n$ already introduced. For example, if we take $N=2$
in eqns. (\ref{tauappr}) and (\ref{tauappr2}), and make use of the following
central difference approximations for the second and fourth derivative of
$\delta(x)$,
\be
{{d^2\delta}\over{dx^2}}={{\delta(x+d)-2\delta(x)+\delta(x-d)}\over{d^2}},
\label{centdiff2}
\ee
\be
{{d^4\delta}\over{dx^4}}={{\delta(x+2d)-4\delta(x+d)+6\delta(x)
-4\delta(x-d)+\delta(x-2d)}\over{d^4}},
\label{centdiff4}
\ee
then we can solve for the coefficients $c_2$ and $c_4$.  In this case, 
these are found to be 
\be
c_2=-d^2(a_1+a_2),
\label{c2}
\ee
\be
c_4=-d^4a_2.
\label{c4}
\ee
This exercise highlights the 
parallels between our non-local approach and recent gradient 
correction models (e.g.[\ref{fleck}], [\ref{aif}]), although in this 
paper we will not explore the
gradient correction form of these equations further.

\subsection{ Energy of Crystalline Slip Within the Non-Local Formulation.}

The simplest test of the non-local model is to re-examine the energetics
of idealized
slip distributions as originally presented in fig.~(\ref{singss}).
Recall that in this figure the 
predictions of the local model are given by the constant dashed
line near the top of each graph, whereas the data points are exact calculations
of the misfit energy as determined using atomistics.	The solid curves
are the results of using the non-local model, which are
in significantly better agreement with the exact results than are
the local results.  In each case, the energetics using the non-local
model exhibit
the correct trend as a function of slip gradients, unlike the
original model which does not capture gradient effects.	
The non-local model
is not able to capture the anomalous upturn in the atomistic results for the
$\{111\}[110]$ screw orientation, a pathology of this particular slip 
deformation which results from putting atoms into highly unfavourable 
proximity across the slip plane. Nonetheless from the standpoint of purely
energetic considerations, the non-local model shows a marked
improvement over the local model, without any great cost in
model complexity.  

Emboldened by our observations that the non-local
correction improves the energetic description of crystalline slip, we now
proceed to a result of greater interest, namely, the
effect of the non-local terms in the context of realistic 
dislocation core structures.

\section{The Non-local Model of Realistic Dislocation Cores}

As an example, we apply the non-local model to the
determination of the core structure for a straight dislocation with
a planar core.	For this purpose we consider the Lomer dislocation in fcc
Al, with a $<110>$ line direction and
Burgers vector ${a\over2}<\bar110>$.  

We obtain the Lomer core structure using two schemes.	The first makes
use of a simplified form for the interplanar potential $\Phi(\delta)$ that
allows for a closed form analytic solution for the slip distribution
in Fourier space.  In the second determination of the Lomer core, the 
full-blown atomistic result for $\Phi(\delta)$ is used and the core structure
is computed numerically.  Each of these results is then compared to the
slip distribution taken directly from the atomic positions resulting from 
full relaxation of the atomistic degrees of freedom.  The first scheme, 
which admits of an analytic solution,	
is of interest as a method for testing the numerical procedures used in 
the second scheme.  At the same time, it demonstrates that 
considerable analytical progress
can be made with the non-local formulation.  The second, fully 
 numerical solution
demonstrates that stable core structures are readily obtainable within
the non-local formulation.  We will see that while the
effect of the non-local terms on the slip distribution and core structure
are subtle to the eye, they represent a significant improvement when
quantified in terms of the predicted misfit energy of the dislocation core.

\subsection{ Eigenstrains solution for $\delta$ within the quadratic well 
approximation.}

In this section, we describe the procedure used in determining an analytic 
result for the core structure of the Lomer dislocation.  The important 
approximation in this procedure in relation to the results of the subsequent
section is the simplified form used for the interplanar potential
$\Phi(\delta)$.

Within the Peierls framework, the determination of $\delta(x)$ is equivalent
to finding the core structure of the dislocation.  Once $\delta(x)$ is
found, the elastic displacement fields away from the slip plane can be
computed from an integral of the Volterra kernel over the entire slip plane.
As well, the misfit and elastic energy are then fully specified by the 
slip distribution.
Therefore, we seek the slip distribution $\delta(x)$ which minimizes the total
energy functional.  The energy associated with the dislocation, 
which is a functional of the slip distribution $\delta(x)$, 
can be expressed as the sum of three parts
\be
E_{tot}=E_L + E_{NL} + E_{B}.
\label{etot}
\ee
$E_L$ is the misfit energy as determined within the traditional
local model and given in eqn.~(\ref{misfitloc}),
$E_{NL}$ is the non-local correction
term added to the misfit energy in eqn.~(\ref{misfitnloc}),
and $E_B$
is the elastic energy of the bulk region.  The elastic term 
is obtained by superimposing
the elastic interaction energy for a distribution of infinitesimal
dislocations with Burgers vector density $-d\delta/dx$,
\be
E_B={{1}\over{2}}\int_{-\infty}^{\infty}\int_{-\infty}^{\infty}
B\log({{R}\over{|x-x'|}}){{d\delta(x)}\over{dx}}
{{d\delta(x')}\over{dx'}}dxdx'.
\label{ebulk}
\ee
In this equation, $R$ is a measure of the size of the bulk region.
Making use of integration by parts, it is possible to eliminate
the constant $R$ from the analysis.  $B$ is defined as
\be
B=2C_{ij}s_is_j
\label{bigb}
\ee
where $s_i$ is the $i$th component of
the slip direction for the dislocation of interest and
$C_{ij}$ is the prelogarithmic energy tensor, discussed in detail, 
for example, in Bacon, Barnett and Scattergood [\ref{bbs}].
In the case of an isotropic solid $B$ reduces to
\be
B_{iso}={{\mu}\over{2 \pi (1-\nu)}}.
\label{bigbiso}
\ee

\begin{figure}
\vspace{0.1in}
\centerline{\psfig{figure=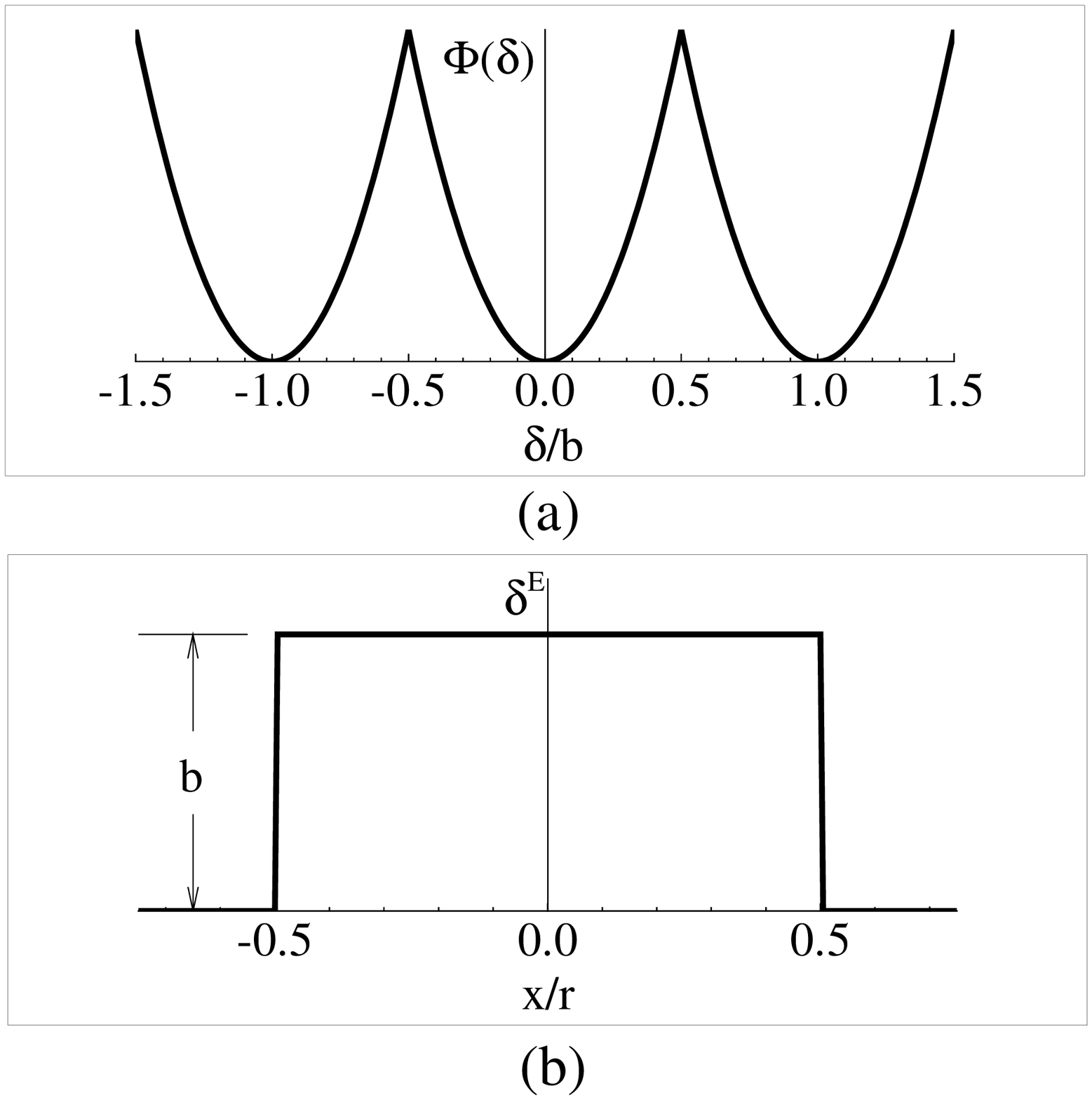,height=3.7truein}}
\caption{(a) Quadratic well approximation to the local interplanar
potential.  (b) Eigenslip, $\delta^E(x)$ for the dipole configuration.}
\label{quadwell}
\end{figure}

\begin{figure}
\vspace{0.1in}
\centerline{\psfig{figure=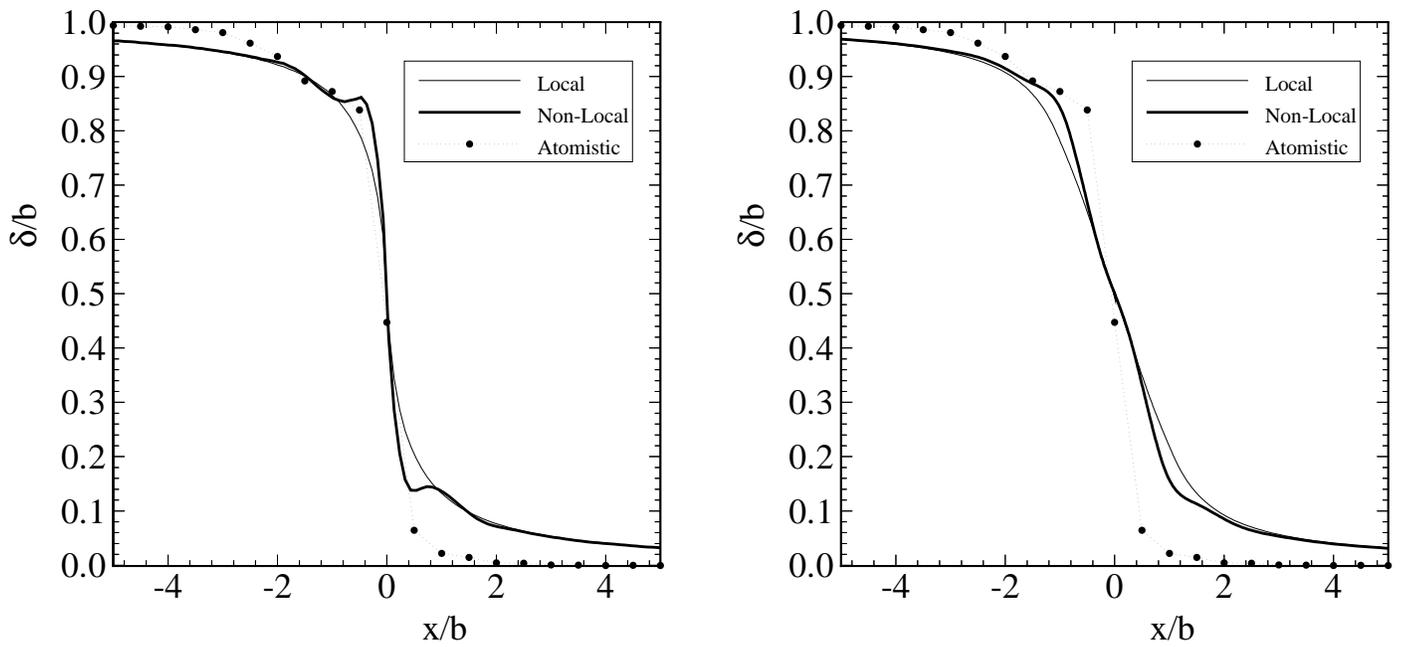,height=3.7truein}}
\caption{Comparison of core structures for the Lomer dislocation in
Al
as obtained from the continuum model with exact atomistic results.  (a)
Structure obtained
using the quadratic well model of the interplanar potential and (b)
with the full numerical solution to the Peierls-Nabarro equation.}
\label{dellom}
\end{figure}

For the purposes of obtaining an analytic solution, we make the approximation
that the local 
interplanar potential $\Phi(\delta)$ takes the form of a periodic
array of quadratic wells as shown in figure (\ref{quadwell}a).  We 
then employ the method of eigenstrains as described by Mura [\ref{mura}] 
to minimize the energy functional
with respect to the slip distribution for a dislocation dipole with spacing
$r$.  The dipole configuration is used so that the slip distribution
goes to zero at $x=\pm \infty$.  By making $r$ large, the dipole 
becomes two isolated dislocations of opposite sign.  
This approach allows us to rewrite the local contribution to the
energy as
\be
E_L=\int_{-\infty}^{\infty}c(\delta(x)-\delta^E(x))^2dx.
\label{esubp}
\ee
In this equation the constant $c$ is determined by fitting the
quadratic wells to linear elasticity at small values of $\delta-\delta^E$.
This leads to $c=\mu/2a$
where $a$ is the spacing between slip planes and
$\mu$ is the relevant shear modulus for the slip system of interest.
$\delta^E$ in this expression is the ``eigenslip'' which ensures that the 
dipole slip configuration is enforced during the energy minimization.
$\delta^E(x)$ is
shown in figure (\ref{quadwell}b), where $b$ is the Burgers vector
and $r$ is the dipole spacing.	We will need
the Fourier transform of this eigenslip, which is found to be
\be
 \hat{\delta}^E=\int_{-r/2}^{r/2}be^{-ikx}dx
={{2b}\over{k}}\sin{{kr}\over2}.
\label{eigenfour}
\ee
Making use of 
eqn.~(\ref{funtrans}), we can replace $\delta$, $\delta^E$ and $K$ with their 
Fourier transforms, allowing for a Fourier space representation of the total
energy.  Integrating by parts on the bulk energy terms and assuming that the
order of integrations can be switched allows us to simplify the
energy expression to
\be
E_{tot}=\int_{-\infty}^{\infty}\Biggl[ 
\hat{\delta}(k)\hat{\delta}(-k)
   \bigl( {c\over{2\pi}}+{{\hat{K}(k)}\over{2\pi}}+{B\over4}|k| \bigr)
-{c\over\pi}\hat{\delta}(k)\hat{\delta}^E(-k)
+{c\over{2\pi}}\hat{\delta}^E(k)\hat{\delta}^E(-k)
\Biggr]dk.
\label{etotfour}
\ee
Note that at this point, our solution is predicated upon a knowledge
of the Fourier space features of the interplanar kernel.
Taking variations with respect to $\hat{\delta}$ and setting 
$\delta E_{tot}=0$ we find
\be
\hat{\delta}(k)={{2b\sin{{kr}\over2}}
  \over{k[1+{{\hat{K}(k)}\over{c}}+{{\pi B}\over{2c}}|k|]}}
\label{deltafour}
\ee
where we have made use of the fact that $\hat{K}(k)$ is an even function and
used eqn.~(\ref{eigenfour}) to replace the Fourier transform
of the eigenslip.	This expression gives us the Fourier transform of the
slip distribution for the dislocation dipole.  Although it cannot be
transformed into real space explicitly, the transformation integral
is straightforward to perform numerically once the form of $\hat{K}$ is
known.  Thus we numerically compute the slip distribution from the 
expression
\be
\delta(x)= {{2b}\over\pi}\int_0^\infty
{{\sin{{kr}\over2}\cos kx}
  \over{k[1+{{\hat{K}(k)}\over{c}}+{{\pi B k}\over{2c}}]}}
dk.
\label{delreal}
\ee
This result is plotted in fig.~(\ref{dellom}) for the case of $\hat{K}=0$
(which corresponds to the local Peierls model solution) and with 
$\hat{K}$ from 
fig.~(\ref{klibrary}) for the $(100)[011]$edge (Lomer) slip system.  
Note that only 
one member of the dipole pair is shown because of the symmetry of 
the distribution.  We see in fig.~(\ref{dellom}a)
that the effect of the non-local correction is to introduce  
small oscillations in the slip distribution and increase its slope slightly 
at the dislocation core.  At first, the oscillations may appear to be
unphysical, but recall that this continuous curve is
really a representation of the slip for a discrete set of lattice sites.
Therefore, only the values at the lattice sites are germane for the
atomic positions implied by the solutions.
It is interesting to note that this 
analytic model provides a reasonable approximation to the dislocation
core, although the differences between the local and non-local results
are hard to quantify.
We should not expect the continuum model to accurately capture
subtle details of the atomic core, but it is encouraging that the model does
lead to a stable core configuration that is similar to the exact result.

\subsection{ Numerical solution for the slip distribution.}

It is now of interest to see what effect the non-local
correction has on the dislocation core
structure when the correct atomistically
obtained $\Phi(\delta)$ is used.  In this case, it is
necessary to resort to a full numerical solution of the governing 
equation.

The numerical approach for finding $\delta(x)$ follows closely the 
approach of Beltz and Freund [\ref{freund}], to which we refer the 
interested reader for more details.  We begin with the well known 
Peierls-Nabarro equation [\ref{hirth}]
\be
\tau[\delta(x)] = -{{\mu}\over{2\pi(1-\nu)}}
\int_{-\infty}^{\infty}{{d\delta(x')/dx'}\over{x-x'}}dx',
\label{pneqn}
\ee
into which we can substitute eqn.~(\ref{tauappr}) for $\tau[\delta(x)]$.  
By an appropriate change of variables, the domain of integration can
be collapsed onto the
finite domain (-1,1).	This domain can be discretized and
the integration carried out by making use of the Gauss-Chebyshev integration 
techniques described in Erdogan and Gupta [\ref{gupta}].  This approach 
reduces the problem to a set of nonlinear 
algebraic equations which can be solved iteratively via the Newton-Raphson
method, leading to a discrete representation of the slip distribution
$\delta(x)$.  The results of such a calculation for the Lomer core structure
are presented in fig.~(\ref{dellom}b).  Again, we note that the resulting 
core structure is not changed dramatically relative to the local model
 as a result of
adding the non-local
correction.  Neither the local nor the non-local model can fully capture
the details of the core, and it is difficult, on the basis of visual 
inspection, to quantify the
virtues of including the non-local correction when considering only
the slip distributions.

The fact that the structural differences between the local and non-local
models are small does not imply that the non-local effect is 
unimportant.  Instead, the important quantity to compare between the 
two models is their predictions for the total energy of the dislocation.
The way the elastic energy is computed for the two models is identical, and
the highly similar predicted cores mean that the value of the elastic
energy for the two models will be about the same.  On the other hand, 
the misfit energy is computed differently, and even identical core 
structures will lead to different misfit energies by virtue of the non-local
energy term.

Using the method described in Miller and Phillips [\ref{miller}] to isolate
the misfit energy from the strain energy in the bulk regions, we can compute
the exact atomistic misfit energy of the lomer dislocation to
be $0.1362 eV/\AA^2$.	This is
accomplished by representing the atoms in the upper and lower bulk regions
by nodes in a finite element mesh, and computing the strains (and
consequent strain energy) in this mesh via the Cauchy-Born rule 
[\ref{ericksen}].
On the other hand, we can compute this energy using the local and non-local
models from eqns.~(\ref{misfitloc}) and (\ref{misfitappr}) respectively.
The local model predicts an energy of $0.1979 eV/\AA^2$,
about $45\%$ greater than the exact energy.  The non-local result is 
$0.1471 eV/\AA^2$,
only $8\%$ greater than the atomistic result.  This result demonstrates
that although the core structures in the local and non-local formulations
are for practical purposes indistinguishable, the non-local treatment
of the core energies is significantly more accurate.  We would expect 
similar adjustments to the cohesive zone model estimate of the Peierls
stress.

\section{Conclusions}

Motivated by previous work which demonstrated a failure of the 
local Peierls framework to accurately describe the energy of interplanar slip
for the types of slip distributions found in real dislocation cores, we
proposed a non-local formulation of this framework.  We then proceeded to
outline a set of simple atomistic calculations whereby one can obtain
the non-local interplanar kernel required in the formulation of the 
model.  To demonstrate that the model improved estimates of 
the slip energy, we compared these results to purely atomistic calculations 
and showed that the non-local model improves the agreement between 
the cohesive zone model energies and explicit atomistic energies.

Given that the non-local model leads to better energetic descriptions of slip
distributions, we proceeded to demonstrate the model by computing the
structure of a straight dislocation with a planar core, namely,
 the Lomer dislocation
in Al.  It was found that the differences between the local and non-local
results, and between either of them and the atomistic core, were subtle
and difficult to quantify when considering only the spatial structure
of the dislocation.  However, the important quantitative measure
of the performance of the models is the energy, which the 
local model predicted to be $45\%$ larger than the exact atomistic 
result.  The non-local model significantly improved the energetic
description of the core, overestimating the exact result by only $8\%$.
Nevertheless, the negligible changes in the non-local description
of the core structure serve as a reminder that the cohesive zone approach
appears to lack the flexibility to really serve as a generic basis for 
mixed atomistic and continuum studies of dislocations.  Even in its
non-local form, this framework restricts the slip to a particular slip plane,
thus forbidding the emergence of complex cores such as those found in bcc
metals.

Future work in the context of cohesive zone models 
could include an attempt to integrate non-local effects
into other models of the mechanical behaviour of materials. One 
example of interest is the model of Rice [\ref{rice}] and the
numerous related works that describe dislocation emission from crack
tips.  Preliminary investigations in this area suggest that the
concepts outlined here can be used for such problems [\ref{miller2}].  
The basic
equation to be solved in the case of a dislocation near a crack tip
is a modified version of the Peierls-Nabarro equation with additional 
terms due to the presence of the crack.  Any realistic solution of this
equation requires a numerical procedure, and the simple form
of the non-local corrections in eqns. (\ref{misfitappr}) and (\ref{tauappr})
mean that the additional computational cost in the non-local formulation
is small.
The most serious obstacle in this case seems to be the proper treatment of
the boundary conditions in the non-local setting, and most importantly
the stress free boundaries at the crack faces.  However, the
fact that non-locality plays an important role at the atomic level
means that a correct non-local treatment of the atomistically sharp crack
may be an important contribution to our understanding of crack tip
phenomena.

Another possible direction for this work is the integration of the 
non-local formulation into the recently proposed ``semi-discrete Peierls 
framework'' of Bulatov [\ref{bulatov}].  Results using this version of 
the Peierls model show significant improvement to the conventional
Peierls framework, but 
some error remains.  It is possible that including non-local effects into
this model may further improve its agreement with atomistics while 
retaining its tractability.

A third future direction for this research should be an effort to
compute other non-local kernels analogous to $K(x-x')$.
For example, it may be possible to directly compute from atomistics
the non-local elastic moduli
introduced by Eringen and co-workers [\ref{eringen1}]--[\ref{eringen3}] 
for bulk crystals.  This would
eliminate guesswork about their appropriate form
and base them solidly on their atomistic underpinnings.  Other recent
work ([\ref{tad1}],[\ref{tad2}]) substantiates our belief in the critical
role played by constitutive non-locality in the description of atomic scale
defects and calls for continued efforts to put such models on a clear
analytic footing.

\section*{Acknowledgements}

This work was supported by the Natural Sciences and Engineering
Research Council of Canada, by NSF grants CMS-9414648 and CMS-9502020
and by the National Science Foundation under the Materials Research
Group grant No. DMR-9223683.  We are grateful to S. Foiles and M. Daw
for the use of their code Dynamo.  Finally, it is a pleasure to
acknowledge useful conversations with V. Bulatov, A. Carlsson,
M.S. Duesbery, J.R. Rice, V. Shenoy and E. Tadmor.

\section*{References}

\begin{enumerate}
\setlength{\itemsep}{0.ex}

\item Peierls, R.E., 1940, {\it Proc. Phys. Soc. Lond.,} {\bf 52}, 34.
\label{peierls}

\item Barenblatt, G.I., 1962, {\it Adv. Appl. Mech.} {\bf 7}, 55.
\label{baren}

\item Dugdale, D.S., 1959, {\it J. Mech. Phys. Sol.,} {\bf 8}, 100.
\label{dugdale}

\item Bilby, B.A., A.H. Cottrell and K.H. Swinden, 1962, 
{\it Proc. Roy. Soc. Lond.,} {\bf A272}, 304.
\label{bcs}

\item Miller, R. and R. Phillips, 1996, {\it Phil. Mag. A,} {\bf 73}, 
 803. 
\label{miller}

\item Rice, J.R., 1992, {\it J. Mech. Phys. Sol.,} {\bf 40}, 239.
\label{rice}

\item Duesbery, M.S., R. Taylor and H.R. Glyde, 1973, {\it Phys. Rev. B}, 
{\bf 8}, 1372.
\label{vibes}

\item DiVincenzo, D.P., 1986, {\it Phys. Rev. B.}, {\bf 34}, 5450.
\label{vibes2}

\item Fleck, N.A., G.M. Muller, M.F. Ashby and J.W.Hutchinson, 1994,
{\it Acta Met. et Mat.,} {\bf 42}, 475.
\label{fleck}

\item Aifantis, E.C., 1992, {\it Int. J. Engng. Sci.}, 
{\bf 30}, 1279.
\label{aif}

\item Eringen, A. C., C.G. Speziale and B.S. Kim, 1977, {\it J. Mech. Phys. 
Sol.,} {\bf 25}, 339.
\label{eringen1}

\item Eringen, A. C. and F. Balta, 1979, {\it Crystal Lattice Defects,}
 {\bf 8}, 73.
\label{eringen2}

\item Eringen, A. C., 1987, {\it Res. Mech.,} {\bf 21}, 313.
\label{eringen3}

\item Hirth, J.P., and J. Lothe, 1992, {\it Theory of Dislocations} 
(Malabar, Florida: Krieger).
\label{hirth}

\item Foreman, A.J., M.A. Jaswon and J.K. Wood, 1951, 
{\it Proc. Phys. Soc. Lond. A,} {\bf 64}, 156.
\label{foreman}

\item Sun, Y., G. E. Beltz and J. R. Rice, 1993, {\it Mat. Sci. Eng. A,} 
{\bf 170}, 67.
\label{ricebs}

\item Kaxiras, E. and M. S. Duesbery, 1993, {\it Phys. Rev. Lett.,} 
{\bf 70}, 3752.
\label{kax}

\item Daw, M.S. and M. I. Baskes, 1984, {\it Phys. Rev. B,} {\bf 29}, 6443.
\label{eam}

\item Ercolessi, F. and J. B. Adams, 1994, {\it Europhys. Lett.}
{\bf 26}, 583.
\label{ercadams}

\item Bacon, D.J., D.M. Barnett and R.O. Scattergood, 1979, 
{\it Prog. Mat. Sci.,} {\bf 23}, 51.
\label{bbs}
 
\item Mura, T., 1984, {\it Micromechanics of Defects in Solids}, 2nd Ed., 
(The Netherlands: Kluwer Academic).
\label{mura}

\item Beltz, G.E., and L.B. Freund, 1994, 
{\it Phil. Mag. A,} {\bf 69}, 183.
\label{freund}

\item Erdogan, F., and G.D. Gupta, 1972, 
{\it Q. Appl. Math.,} {\bf 29}, 525.
\label{gupta}

\item Ericksen, J.L., 1984, {\it Phase Transformations and Material 
Instabilities in Solids}, edited by M. Gurtin, 
(New York: Academic Press), pp. 61-77.
\label {ericksen}

\item Miller, R., R. Phillips, G. Beltz and M. Ortiz, 1996, unpublished.
\label {miller2}

\item Bulatov, V. and E. Kaxiras, 1996, to be submitted 
to {\it Phys. Rev. Lett.}
\label {bulatov}

\item Tadmor E.B., M. Ortiz and R. Phillips, 1994, {\it Phil. Mag. A,} 
{\bf 73}, pp 1529.
\label{tad1}

\item Tadmor, E.B., R. Phillips and M. Ortiz, 1996, {\it Langmuir}, 
{\bf 12}, pp 4529.
\label{tad2}

\end{enumerate}

\end{document}